\documentclass[twocolumn]{aastex631}

\usepackage{graphicx}
\usepackage{xspace}
\usepackage{multirow}

\newcommand{\chimefrb}{CHIME/FRB\xspace}
\newcommand{\chimeslow}{CHIME/Slow\xspace}
\newcommand{\presto}{\texttt{PRESTO}\xspace}
\newcommand{\hdbscan}{\texttt{HDBSCAN}\xspace}
\newcommand{\fetch}{\texttt{FETCH}\xspace}

\newcommand{\dm}{\ensuremath{\mathrm{pc~cm}^{-3}}\xspace}

\begin{document}

\title{\chimeslow overview and pilot survey: A new backend to search for second-duration radio transients with the CHIME telescope}

\correspondingauthor{Sujay Mate}
\email{sujay.mate@rrimail.rri.res.in, sujay.mate@gmail.com}

\author[0000-0001-5536-4635]{Sujay Mate}
\affiliation{Department of Astronomy and Astrophysics, Tata Institute of Fundamental Research, Mumbai, 400005, India}
\affiliation{Raman Research Institute, C. V. Raman Avenue, Sadashivanagar, Bangalore, Karnataka, 560080, India}

\author[0000-0003-4443-4508]{Kevin Luke}
\affiliation{LUX, Observatoire de Paris, Université PSL, Sorbonne Université, CNRS, 92190 Meudon, France}

\author[0000-0002-5342-163X]{Yash Bhusare}
\affiliation{National Centre for Radio Astrophysics, Post Bag 3, Ganeshkhind, Pune, 411007, India}

\author[0000-0003-0477-7645]{Arvind Balasubramanian}
\affiliation{Department of Astronomy and Astrophysics, Tata Institute of Fundamental Research, Mumbai, 400005, India}
\affiliation{Indian Institute of Astrophysics, Koramangala II Block, Bangalore 560034, India}

\author[0000-0002-4795-697X]{Ziggy Pleunis}
\affiliation{Anton Pannekoek Institute for Astronomy, University of Amsterdam, Science Park 904, 1098 XH Amsterdam, The Netherlands}
\affiliation{ASTRON, the Netherlands Institute for Radio Astronomy, Oude Hoogeveensedijk 4, 7991 PD Dwingeloo, The Netherlands}

\author[0000-0002-7374-7119]{Paul Scholz}
\affiliation{Department of Physics and Astronomy, York University, 4700 Keele Street, Toronto, ON MJ3 1P3, Canada}

\author[0000-0003-2548-2926]{Shriharsh P. Tendulkar}
\affiliation{Department of Astronomy and Astrophysics, Tata Institute of Fundamental Research, Mumbai, 400005, India}
\affiliation{National Centre for Radio Astrophysics, Post Bag 3, Ganeshkhind, Pune, 411007, India}

\author[0000-0002-3615-3514]{Mohit Bhardwaj}
\affiliation{Department of Space Planetary Astronomical Sciences and Engineering, Indian Institute of Technology KanpuR, Uttar Pradesh, India}

\author[0000-0002-1800-8233]{Charanjot Brar}
\affiliation{NRC Herzberg Astronomy and Astrophysics, 5071 West Saanich Road, Victoria, BC V9E2E7, Canada}

\author[0000-0003-4098-5222]{Fengqiu Adam Dong}
\affiliation{Department of Physics and Astronomy, York University, 4700 Keele Street, Toronto, ON MJ3 1P3, Canada}
\affiliation{National Radio Astronomy Observatory, 520 Edgemont Rd, Charlottesville, VA 22903, USA}

\author[0000-0001-8384-5049]{Emmanuel Fonseca}
\affiliation{Department of Physics and Astronomy, West Virginia University, PO Box 6315, Morgantown, WV 26506, USA}
\affiliation{Center for Gravitational Waves and Cosmology, West Virginia University, Chestnut Ridge Research Building, Morgantown, WV 26505, USA}

\author[0000-0002-3382-9558]{B. M. Gaensler}
\affiliation{Department of Astronomy and Astrophysics, University of California, Santa Cruz, 1156 High Street, Santa Cruz, CA 95060, USA}
\affiliation{Dunlap Institute for Astronomy and Astrophysics, 50 St. George Street, University of Toronto, ON M5S 3H4, Canada}
\affiliation{David A. Dunlap Department of Astronomy and Astrophysics, 50 St. George Street, University of Toronto, ON M5S 3H4, Canada}

\author[0000-0003-2317-1446]{Jason Hessels}
\affiliation{Anton Pannekoek Institute for Astronomy, University of Amsterdam, Science Park 904, 1098 XH Amsterdam, The Netherlands}
\affiliation{ASTRON, the Netherlands Institute for Radio Astronomy, Oude Hoogeveensedijk 4, 7991 PD Dwingeloo, The Netherlands}
\affiliation{Department of Physics, McGill University, 3600 rue University, Montr\'eal, QC H3A 2T8, Canada}
\affiliation{Trottier Space Institute, McGill University, 3550 rue University, Montr\'eal, QC H3A 2A7, Canada}

\author[0000-0002-8043-0048]{Jeff Huang}
\affiliation{Department of Physics, McGill University, 3600 rue University, Montr\'eal, QC H3A 2T8, Canada}
\affiliation{Trottier Space Institute, McGill University, 3550 rue University, Montr\'eal, QC H3A 2A7, Canada}

\author[0009-0009-0938-1595]{Naman Jain}
\affiliation{Department of Physics, McGill University, 3600 rue University, Montr\'eal, QC H3A 2T8, Canada}
\affiliation{Trottier Space Institute, McGill University, 3550 rue University, Montr\'eal, QC H3A 2A7, Canada}

\author[0000-0003-3457-4670]{Ronniy C. Joseph}
\affiliation{Department of Physics, McGill University, 3600 rue University, Montr\'eal, QC H3A 2T8, Canada}
\affiliation{Trottier Space Institute, McGill University, 3550 rue University, Montr\'eal, QC H3A 2A7, Canada}

\author[0000-0001-9345-0307]{Victoria M. Kaspi}
\affiliation{Department of Physics, McGill University, 3600 rue University, Montr\'eal, QC H3A 2T8, Canada}
\affiliation{Trottier Space Institute, McGill University, 3550 rue University, Montr\'eal, QC H3A 2A7, Canada}

\author[0009-0004-4176-0062]{Afrokk Khan}
\affiliation{Department of Physics, McGill University, 3600 rue University, Montr\'eal, QC H3A 2T8, Canada}
\affiliation{Trottier Space Institute, McGill University, 3550 rue University, Montr\'eal, QC H3A 2A7, Canada}

\author[0000-0002-7164-9507]{Robert Main}
\affiliation{Department of Physics, McGill University, 3600 rue University, Montr\'eal, QC H3A 2T8, Canada}
\affiliation{Trottier Space Institute, McGill University, 3550 rue University, Montr\'eal, QC H3A 2A7, Canada}

\author[0000-0001-8845-1225]{Bradley W. Meyers}
\affiliation{Australian SKA Regional Centre (AusSRC), Curtin University, Kent Street, Bentley, WA 6102, Australia}
\affiliation{International Centre for Radio Astronomy Research (ICRAR), Curtin University, Kent Street, Bentley, WA 6102, Australia}

\author[0000-0001-8292-0051]{Nikola Milutinovic}
\affiliation{Department of Physics and Astronomy, University of British Columbia, 6224 Agricultural Road, Vancouver, BC V6T 1Z1 Canada}

\author[0000-0003-0510-0740]{Kenzie Nimmo}
\affiliation{Center for Interdisciplinary Exploration and Research in Astronomy, Northwestern University, 1800 Sherman Avenue, Evanston, IL 60201, USA }

\author[0000-0002-6823-2073]{Kaitlyn Shin}
\affiliation{Division of Physics, Mathematics, and Astronomy, California Institute of Technology, Pasadena, CA 91125, USA}

\author[0009-0003-6054-8035]{David Spear}
\affiliation{Department of Physics and Astronomy, University of British Columbia, 6224 Agricultural Road, Vancouver, BC V6T 1Z1 Canada}

\author[0000-0001-9784-8670]{Ingrid Stairs}
\affiliation{Department of Physics and Astronomy, University of British Columbia, 6224 Agricultural Road, Vancouver, BC V6T 1Z1 Canada}

\author[0000-0001-7509-0117]{Chia Min Tan}
\affiliation{International Centre for Radio Astronomy Research (ICRAR), Curtin University, Kent Street, Bentley, WA 6102, Australia}



\begin{abstract}
We present an overview of \chimeslow, a real-time transient search backend under development to search for second-duration radio transients using the CHIME telescope, and results obtained from a pilot survey carried out using the prototype version of the search pipeline. The prototype \chimeslow pipeline was tested on archival data obtained in December 2022, January 2023 and February 2023 with a total on-sky time of 17 days with an instantaneous field of view (FoV) of $\sim13$~deg$^2$. In this pilot survey, we detected nine bursts, one from a new non-repeating source, FRB~20230204C, and eight from the known hyperactive repeating source FRB~20220912A. Out of these nine bursts, two bursts from the repeater were not detected by \chimefrb, while the non-repeater was detected in the side-lobe of a beam in \chimefrb, exhibiting shorter burst width and narrower bandwidth compared to the \chimeslow detection. Here we report properties of the bursts, discuss the sensitivity and completeness of the current version of the \chimeslow pipeline, and outline future development to improve its performance. Finally, based on these results, we report the all-sky rate (95\% credible region) of radio transients with burst widths between 16~ms to 5~s, fluence above $5$~Jy~ms and observing frequency of 600~MHz to be between 184 and 4556 bursts sky$^{-1}$ day$^{-1}$.
\end{abstract}



\section{Introduction} \label{sec:intro}
The construction of radio interferometers with large mapping speeds (combining field of view and sensitivity) has led to a proliferation of radio transient detections in recent years. The detection of a large population of fast radio bursts (FRBs) through untargeted surveys (e.g.~\citet{CHIMEFRB2021}) has led to many insights into this mysterious source class, but most instruments lack sensitivity to the widest and most highly scattered (timescales $\gtrsim 100$\,ms) bursts. 

\begin{figure*}
    \centering
    \includegraphics[width=0.8\linewidth]{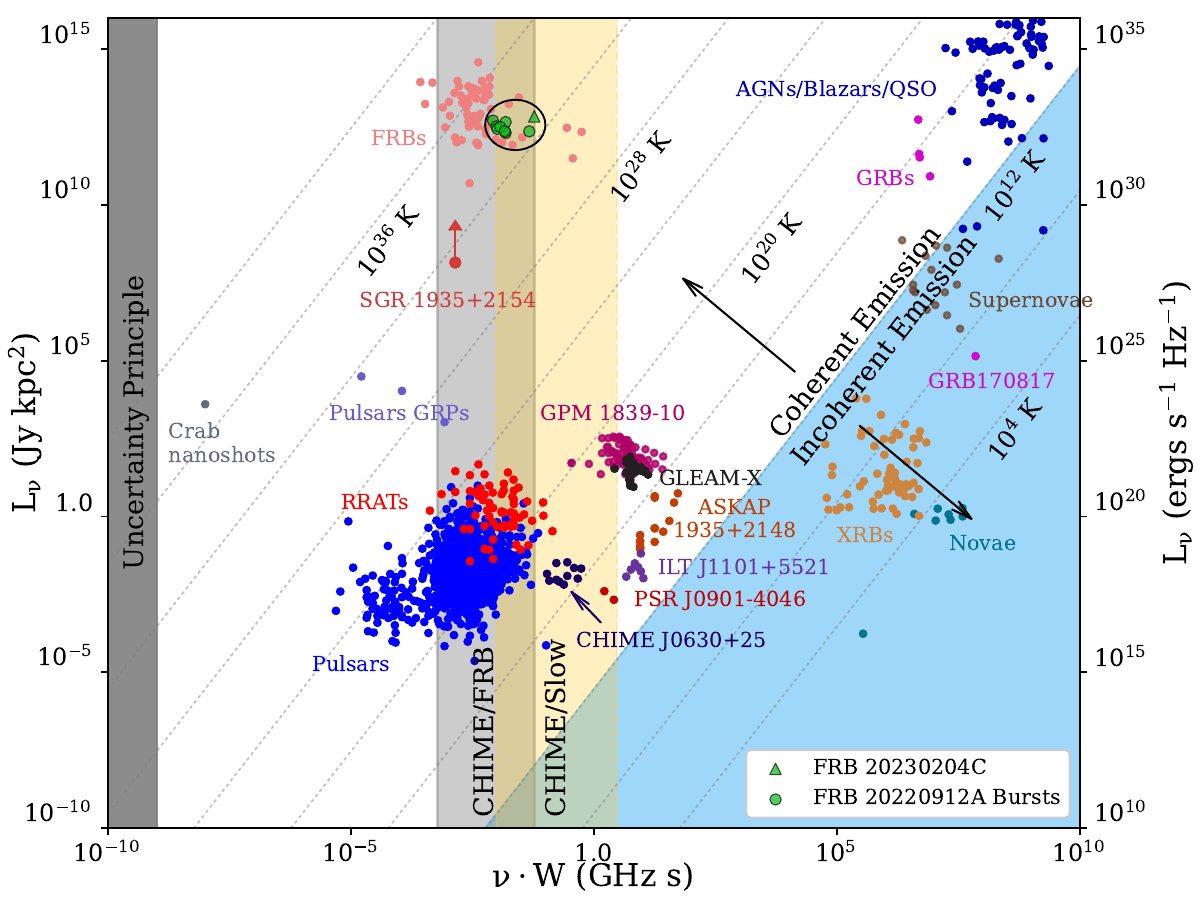}
    \caption{Radio transients associated with compact objects in the pseudo-luminosity vs. pseudo pulse width parameter space. The parts of the parameter space covered by \chimefrb and \chimeslow are highlighted in light gray and light yellow colors, respectively. The green points near the FRB population show the \chimeslow detections from the pilot survey with the circles indicating bursts from FRB~20220912A and the triangle indicating the non-repeating burst FRB~20230204C (See Section~\ref{sec:survey} for more details). Figure adapted from~\citet{Cordes2004} and~\citet{Keane2018} with data from~\citet{Hurley-Walker2022,Hurley-Walker2023,Caleb2022,Caleb2024,Dong2024,de_Ruiter2024} and this work. Note that the shaded regions only mark the widths covered.}\label{fig:transient_ps}
\end{figure*}

Radio transients associated with compact objects span a broad range in the burst width -- luminosity phase space (Figure~\ref{fig:transient_ps}). FRBs and pulsars lie in the short-duration ($\sim$ms) region of the phase space, with FRBs occupying the extreme luminosity end. At much longer timescales ($\sim$minutes to days), emissions associated with active galactic nuclei (AGNs), gamma-ray bursts (GRBs) and X-ray binaries (XRBs) are observed. However, only a handful of transients have been observed at $\sim$seconds duration, leaving a large gap in phase space. One of the reasons for a dearth of second timescale transients has been primarily due to two technological limitations: the lack of gain stability of radio telescopes at these timescales for time-domain searches due to radio frequency interference (RFI) and bright sources transiting through sidelobes, and the computational challenge of doing image plane searches at very high (seconds) time resolution. However, recent advances in technology have allowed telescopes like the MeerKAT, Murchison Widefield Array (MWA), Canadian Hydrogen Intensity Mapping Experiment (CHIME), Australian Square Kilometre Array Pathfinder (ASKAP) and the LOw Frequency ARray (LOFAR) to detect second-duration radio transients over the past few years. These mostly include long-period pulsar-like transients (LPTs) with pulse widths ranging from few hundreds of milliseconds to few tens of seconds~\citep[e.g. see][]{Hurley-Walker2022,Hurley-Walker2023,Caleb2022,Caleb2024,Dong2024,de_Ruiter2024}\footnote{Full list of all-known LPTs with references can be found here: \url{https://lpt.mwa-image-plane.cloud.edu.au/published/tables/1}}.
The sources of these transients are still unknown, with some observations suggesting a neutron star/magnetar origin while other systems seem to be comprised of magnetized white dwarfs or interacting white dwarfs in binaries. Detailed follow-up observations and more detections of such LPTs are needed to understand this new class of sources.

Apart from these detections, some theories propose that compact binary mergers involving at least one neutron star could produce FRB-like flashes that could last for a few hundred milliseconds to a few seconds~\citep[e.g. see][]{Rowlinson2019b,Sridhar2021,Lyutikov2024}. Furthermore, it has been theorized that classical GRBs could also produce a radio flash of similar duration concurrent with the prompt emission if certain emission pathways dominate the GRB jet~\citep[i.e., if the jet is ``Poynting flux" dominated, see][]{Usov2000}. Several efforts have been made to search for FRB-like emission associated with compact binary mergers and GRBs, but none have resulted in a confirmed detection~\citep{Anderson2018a,Rowlinson2019,Starling2020,Rowlinson2021,Tian2022,Curtin2023,Hennessy2023,Chastain2023,Curtin2024}. However, these results are consistent given the search timescales, sky coverage and sensitivity of these searches, and hence more systematic efforts are warranted. The \chimeslow backend, described below, will systematically search for second-timescale radio transients, and aims to fill the as-yet largely unexplored parameter space.

This article is structured in the following way: Section~\ref{sec:pipeline} describes the \chimeslow pipeline and its major components. Section~\ref{sec:survey} presents the results from the pilot survey carried out using the prototype version of the \chimeslow pipeline. Section~\ref{sec:injection} describes the injection and completeness analysis performed to estimate the sensitivity of the current version of the \chimeslow pipeline, which is also used to estimate the all-sky rate of second-duration radio transients in Section~\ref{sec:rate}. Section~\ref{sec:disc} discusses the implications of these results and future plans for the \chimeslow backend. Finally, Section~\ref{sec:concl} presents the summary and conclusions of this work. Appendix~\ref{sec:appendixA} describes additional details about the all-sky rate estimation.

\section{The \chimeslow backend}\label{sec:pipeline}
The \chimefrb backend~\citep{CHIMEFRB2018} searches for FRBs with burst widths up to 100~ms and DMs up to 13,000~\dm. The upper limits on the burst widths and DMs were decided as a compromise between computational cost and statistical optimization to search the DM and burst width parameter space. Although burst widths up to 100~ms are searched, the completeness of the pipeline drops significantly after $\sim$30~ms~\citep[see Fig. 18 of][]{CHIMEFRB2021}. This loss in sensitivity occurs because of the fact that \chimefrb RFI cleaning algorithms were optimized for short-duration transients and the training data used for the machine learning classifier was biased against the long duration events~\citep{Merryfield2023}. The \chimeslow pipeline was designed by taking cognizance of these biases and correcting for them and, as a proof-of-concept, to search for second-duration transients using CHIME telescope.
\begin{figure*}
    \centering
    \includegraphics[width=0.9\linewidth]{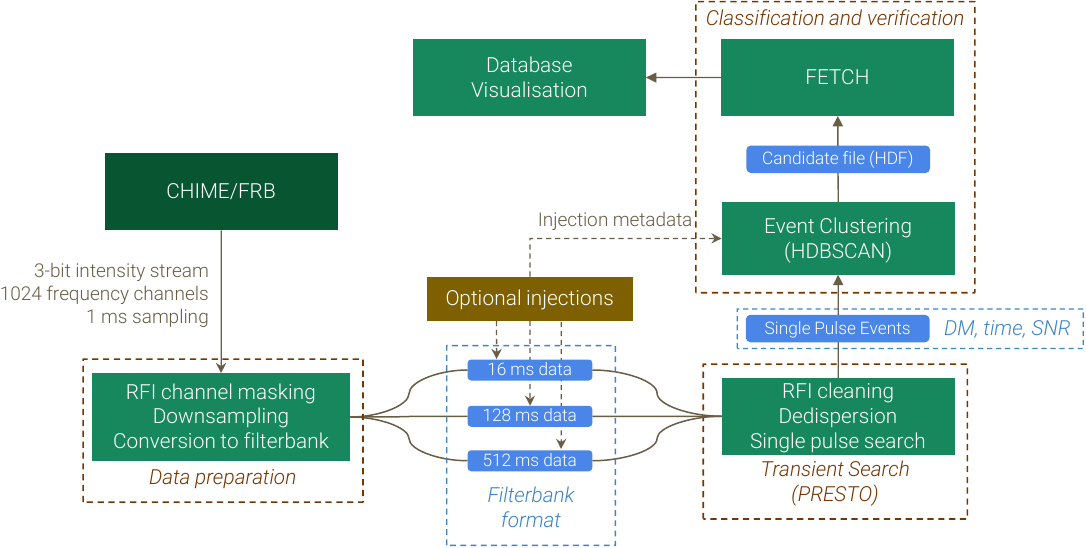}
    \caption{Overview of the \chimeslow pipeline: The diagram shows the major steps in the pipeline and the tools used to perform these steps. The light green blocks represent data processing routines, the blue blocks represent data products that are created during the search and the brown block represent the injection subpackage.}\label{fig:pipeline}
\end{figure*}

\chimeslow uses a $\sim$3-bit Huffman compressed intensity data stream copied from the \chimefrb 8-bit data stream. This encoding was done as part of a development of another telescope backend: the CHIME All-Sky Multiday Pulsar Stacking Search project~\citep[CHAMPSS,][]{CHAMPSS2025}, which searches for pulsars above the period of 10~ms. The factor $\sim$3 reduction in the intensity data (from 8-bit to 3-bit), factor of 16 reduction in frequency resolution (from 16384 channels in case of \chimefrb to 1024 channels in case of \chimeslow) and larger than tens of ms time-sampling necessary to search for longer timescale transients reduces the computational cost of the \chimeslow backend significantly. The downsampling in frequency could impact the search sensitivity for wide low-DM bursts due to intra-channel dispersion smearing. However, other than that, we do not expect any significant loss in sensitivity as the Huffman encoding is lossless and the search is performed at timescales that are comparable to the downsampled time resolution as explained below. The ultimate goal of \chimeslow is to systematically search the sky visible to CHIME for radio transients having pulse widths between 16 ms and 5 s.

The \chimeslow pipeline is based on well-tested tools that include \presto~\citep{Ransom2011}, \hdbscan~\citep{McInnes2017} based clustering and \fetch~\citep{Agarwal2020}. Figure~\ref{fig:pipeline} gives an overview of the pipeline showing all major steps. The 3-bit, 0.98304~ms (referred as 1-ms hereafter) sampled intensity data stream from \chimefrb is downsampled to 16-ms, 128-ms and 512-ms data streams and the search is performed independently on each data stream. The 16-ms data are searched for pulse widths up to 128~ms, the 128-ms data are searched for pulse widths up to 512~ms and the 512-ms data are searched for pulse widths up to 5~s. RFI cleaning is applied at different stages of the pipeline using different algorithms (see Section~\ref{sec:transient_search}) with parameters optimized to search for long-duration transients. These parameters were fixed by carrying out injections of wide duration bursts into test data using the built-in injection routine and recovering them back with the pipeline. The same routine was also used to estimate the completeness and sensitivity of the current version (see Section~\ref{sec:injection}), and will be used to benchmark the real-time system. The following subsections describe the details of each major block (marked with dotted brown rectangles in the Figure~\ref{fig:pipeline}) of the pipeline.

\subsection{Data Preparation}
The first step in the data preparation block is to find and mask RFI channels in the native 3-bit, 1-ms $\times$ 1024 frequency channel intensity data (considered as one `chunk'). The masking is performed by computing kurtosis, skew and standard deviation metrics on the bandpass data and masking the outlier channels. The data are then bandpass corrected and normalized by subtracting the 1~s chunkwise mean to minimize the effect of gain variations\footnote{Similar normalization is also carried out by \chimefrb pipeline.}. These data are downsampled to 16-ms, 128-ms and 512-ms and stored as independent 32-bit filterbank files. The duration of each filterbank file varies based on the downsampling factor to ensure there are enough samples in each file for \presto to perform time-frequency chunk-wise RFI cleaning, dedispersion and single pulse search as explained in the Section~\ref{sec:transient_search}. The 16-ms filterbank file has a duration of $\sim$1~hr, the 128-ms file has a duration of $\sim$2~hrs and the 512-ms file has a duration of $\sim$4~hrs. Between two consecutive files, an overlap of $\sim$128~s is included to ensure no transients are missed at the file boundaries. Along with the filterbank files, the bad channel mask computed at the first step is also stored, which is then passed to successive transient search and classification steps.

\subsection{Transient search}\label{sec:transient_search}
The transient search is performed using \texttt{PRESTO} to search for single pulses in DM-time space. The transient search block consists of three steps: RFI removal, dedispersion, and single pulse search. The following steps are performed independently for the three time-downsampling factors.

\subsubsection{RFI Cleaning}
The narrowband, periodic and impulsive RFI flagging in the time-frequency domain is performed by the \texttt{rfifind} routine of \presto. The input parameters used for the RFI flagging are: \texttt{clip} = 6, \texttt{timesig} = 5, \texttt{freqsig} = 15, \texttt{chanfrac} = 0.5,  \texttt{intfrac} = 0.5 and \texttt{blocks} = 2, 4, 1 for 16-ms, 12-ms, 512-ms time-samplings, respectively. These parameters were qualitatively selected by injecting simulated bursts in test data collected in mid 2022 and maximizing the recovery of the injected bursts after RFI cleaning. A larger injection run was carried out using the standard CHIME/FRB injection routine \citep{Merryfield2023} in August 2025 and the data from this run is being used to optimize the real-time pipeline and measure its selection function. This analysis will be presented in future work.

\subsubsection{Dedispersion}
The data are dedispersed from DM values of 50~\dm to 3000~\dm using the \texttt{prepsubband} routine of \presto. The limits on DM search were set based on the available compute in case of the upper limit, while the lower limit was set in order to be mostly sensitivity to extragalactic transients and avoid Galactic sources\footnote{Note that this choice was made only for the pilot survey, in case of the real-time survey (Section~\ref{sec:disc}), the limits will be revised.}. For 16-ms time-sampling, a DM step of 2~\dm is used, while for 128-ms and 512-ms time-sampling, a DM step of 10~\dm is used. The DM steps were decided using the \texttt{DDPlan} routine of \presto. At this stage, RFI masks computed in the earlier steps (at the data preparation step and by \texttt{rfifind}) are used to mask the noisy channels and time blocks, in addition to the zero-DM subtraction\footnote{We do not perform any detailed investigation of the effect of zero-DM subtraction on the signal. The main aim of the pilot survey was to detect bright events and zero-DM subtraction is effectively mitigating broadband RFI without significantly affecting the SNR for bright bursts. In the real-time pipeline, we are exploring the possibility of using ``flat-fielding'', a technique used by~\cite{Kuiper2026} for the LOFAR telescope. We expect that it will perform better than zero-DM subtraction.} which removes any broadband RFI.

\subsubsection{Single Pulse Search}
The dedispersed time series are searched for single pulses using the \texttt{single\_pulse\_search.py} routine of \presto. Boxcar widths of size 2$^n$ are used for the search with the maximum boxcar width determined by the maximum width (\texttt{-m}) option of the routine. It is set to 0.13~s, 0.6~s and 6~s for the 16-ms, 128-ms, and 512-ms time samplings, respectively. The detection threshold is set at a signal-to-noise (S/N) ratio of 10. The single pulse search outputs a file with the DM, time, S/N and boxcar downfactor (proxy for pulse width) for each detected event. This information is passed to the classification block to identify potential astrophysical events from the RFI.

\subsection{Classification}
The classification block consists of two main steps: clustering of single pulse events in DM-time parameter space, and identification of FRB-like pulses in the clusters classified as potentially astrophysical. The clustering in DM-time space is performed using the \texttt{HDBSCAN} python library. We use \texttt{min\_cluster\_size} and \texttt{min\_samples} equal to 5 and \texttt{cluster\_selection\_epsilon} of 0.1 for this step. These parameters were selected using the injection analysis (see Section~\ref{sec:injection}), and testing the clustering on real data containing pulsar transits. The selected parameters ensure that events that are sparse with a large distance in the phase space ($>$0.1 units) are not clustered together, and only densely populated clusters are selected. For each of the remaining clusters, the event with maximum S/N is passed to the \texttt{FETCH} classification algorithm to identify whether it is an astrophysical FRB-like event or RFI. For each event, \texttt{FETCH} assigns a probability of it being astrophysical in origin. Diagnostic plots are generated for events that have a probability $> 0.7$, as input to the final step of manual classification\footnote{Note that we have not trained \texttt{FECTH} on CHIME data yet, and for this analysis we have used the default models.}.

\section{\chimeslow pilot survey}\label{sec:survey}
A pilot survey was carried out to test the prototype version of the \chimeslow pipeline using the data collected in December 2022, January 2023 and February 2023. These data were collected as part of the development of the CHAMPSS backend. The total on-sky time of these data was 17 days covering $\sim13$~deg$^2$ at a given instance as data were collected for only a fraction of 1024 \chimefrb beams. The total exposure is equivalent to observing the full CHIME sky for one day (see Appendix~\ref{sec:appendixA} for more details on exposure). The data were searched for single pulses in the de-dispersed time series from 16~ms to up to 5~s and for a DM range of 50 to 3000~\dm.
\begin{figure*}[ht]
\centering
\includegraphics[width=\linewidth]{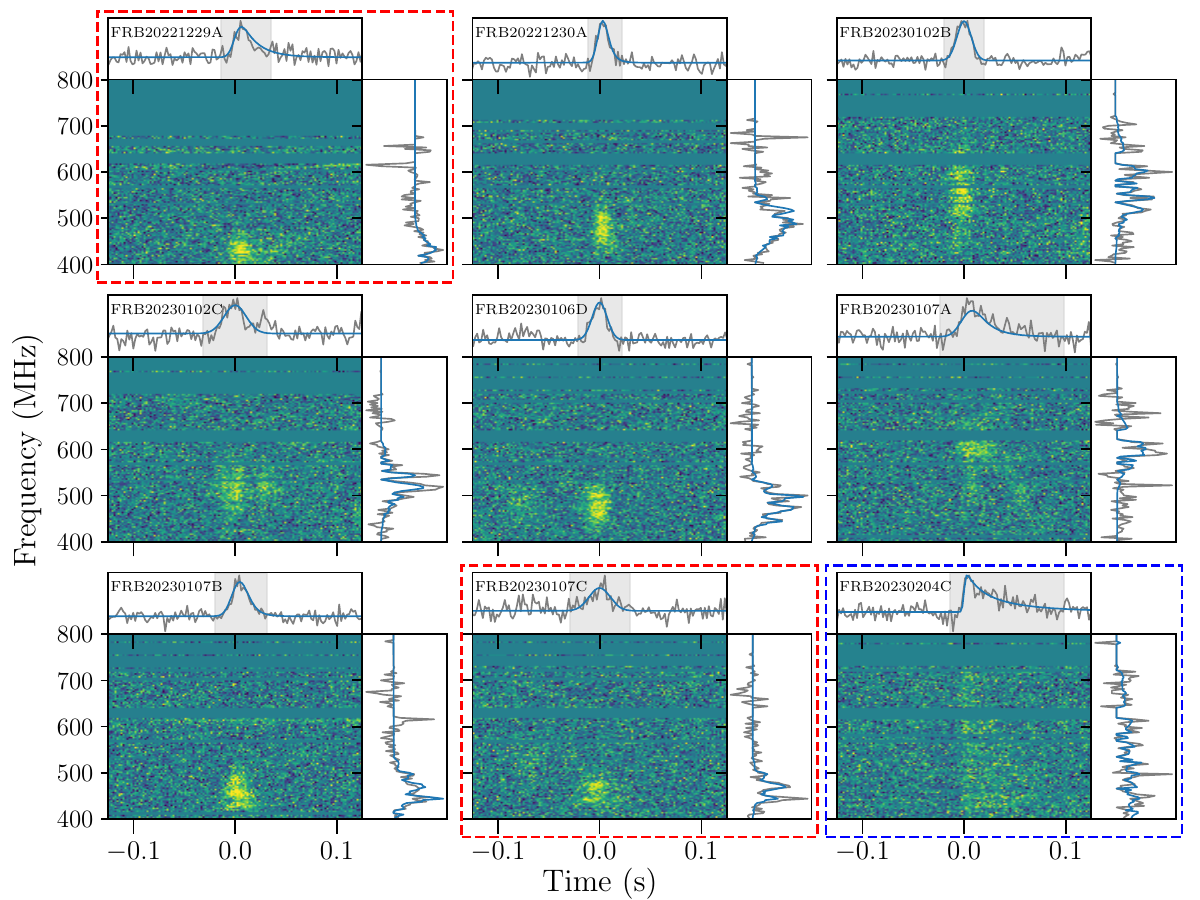}
\caption{Dynamic spectra (``waterfall'' plots), frequency-averaged time series, and time-averaged spectra of bursts detected in the CHIME/Slow pilot survey. The bursts are plotted in order of their arrival times (topocentric) from top-left to bottom-right. Each plot is corrected for the best-fit DM obtained from \texttt{fitburst}. The blue lines in the time-profile and spectrum plot show the best-fit frequency-averaged model for the burst profile and time-averaged spectral model estimated over the burst duration respectively. The last burst, inside the blue box, is the one-off event FRB~20230204C, while all other bursts are from the repeating source FRB~20220912A. Out of these eight bursts, the two marked by red boxes were not detected by CHIME/FRB, while the other six were co-detections.}\label{fig:waterfall}
\end{figure*}

In this survey, we detect nine bursts (Table~\ref{tab:burst_properties}), which are shown in Figure~\ref{fig:waterfall}. Out of these nine bursts, one burst, FRB~20230204C, is from a new FRB source and the other eight bursts are from the hyperactive repeater FRB~20220912A~\citep{McKinven2022,Cook2024}. FRB~20230204C was also detected by CHIME/FRB, but in the side-lobe of neighboring beams, hence exhibiting a narrower bandwidth and pulse width, which enabled its detection as the effective pulse width was in sensitivity range of \chimefrb. Out of the eight FRB~20220912A bursts, six bursts were also detected by CHIME/FRB while two bursts were only detected in CHIME/Slow data.

We model the dynamic spectra of each burst using the \texttt{fitburst}\footnote{\url{https://github.com/CHIMEFRB/fitburst}}~\cite[]{Fonseca2024} package. A burst model is least-squares optimized to find the best-fit DM, intrinsic width, scattering timescale (if any) and spectral properties of the burst. We find that adding an exponential scatter-broadening tail\footnote{We assume scattering index of -4.} to our models improves fits for five out of the nine bursts (for four bursts from FRB~20220912A and FRB~20230204C). We find that FRB~20230204C has an intrinsic width (measured as full-width at half maximum or FWHM) of $3.49\pm1.32$~ms with scattering timescale of $98.31\pm8.33$~ms at 400~MHz, making it one of the most highly scattered FRB detected by CHIME~\citep{CHIMEFRB2026}. For the eight FRB~20220912A bursts, the intrinsic widths are found to be between $10.22 - 25.48$~ms at 400~MHz and the four scattered bursts have scattering timescales between $9.59 - 75.06$~ms at 400~MHz\footnote{400~MHz corresponds to the bottom of the CHIME observation band. The widths and scattering timescales are reported at this frquency as \texttt{fitburst} reports them at the bottom of the band by default.}.

The DM and intrinsic width values obtained from \texttt{fitburst} and the daily calibration metrics stored by \chimefrb are used to determine the peak flux of each burst at 1-ms timescale and the total fluence of each burst. We use the same method as described by~\citet{Andersen2023} for the flux/fluence calibration; and hence it should be noted that the flux and fluence values are lower limits as they are not corrected for the actual burst position. The DM information from \texttt{fitburst} is used to estimate the redshift of FRB~20230204C by solving for the redshift in the Macquart relation~\citep{Macquart2020} using the Bayesian Markov-Chain Monte Carlo (MCMC) simulations described in Appendix A of \citet{Bhardwaj2021b}. For the FRB~20220912A bursts, we use the redshift reported by~\cite{Ravi2023}. The redshift information is used to estimate luminosity distance and the lower limit on the peak luminosity of all bursts

\section{Injections and completeness analysis}\label{sec:injection}
We perform an injection-based completeness analysis to estimate the completeness of \chimeslow pipeline as function of detection S/N and total injected pulse width (intrinsic + scattering). This is carried out independently on the three data streams (16-ms, 128-ms and 512-ms). We inject a total of $\sim$10,000 pulses at each time-sampling in $\sim$4~hr worth of data taken over 10 days and 64 beams. The pulses are injected using the \texttt{simpulse}\footnote{\url{https://github.com/kmsmith137/simpulse}} routine over the entire 400~--~800 MHz band. Table~\ref{tab:injection_params} gives the details of injected parameters including the distribution used to sample each parameter. While injecting, we randomize the beam number to sample different directions in the sky. We estimate the completeness at the end of the clustering step by matching the injected pulses with the detected candidates. A candidate is considered recovered when a cluster coincides with the injected pulse in the DM-time parameter space and has a maximum S/N $\geq$ 10. The details about estimating the completeness profile are described below.

We bin all injections into total injected pulse width space with 10~ms, 20~ms and 500~ms binwidths for pulses injected into 16-ms, 128-ms and 512-ms data respectively. We evaluated the completeness for each of these bins independently. For each bin, we first bin the data into injected S/N space with width of 0.5 S/N units. Ideally, the detected S/N for each of the injected bin should be close to the injected S/N. Therefore, finding the ratio of total detections of injections ($N_{det}$) to the total injections ($N_{inj}$) in each injected S/N bin should give the completeness profile as a function of detected S/N. However, due to the varying RFI and noise conditions in the data and differences in the total pulse width of each pulse, the detected S/N and injected S/N do not have one-to-one mapping. To account for this, the corresponding detected S/N in each injected S/N bin is determined by taking the median of the detected S/N values for all injections in that bin. The completeness is defined as the fraction $N_{det} / N_{inj}$ for the each of the detected S/N bin defined above.
\begin{figure}[h]
\centering
\includegraphics[width=\linewidth]{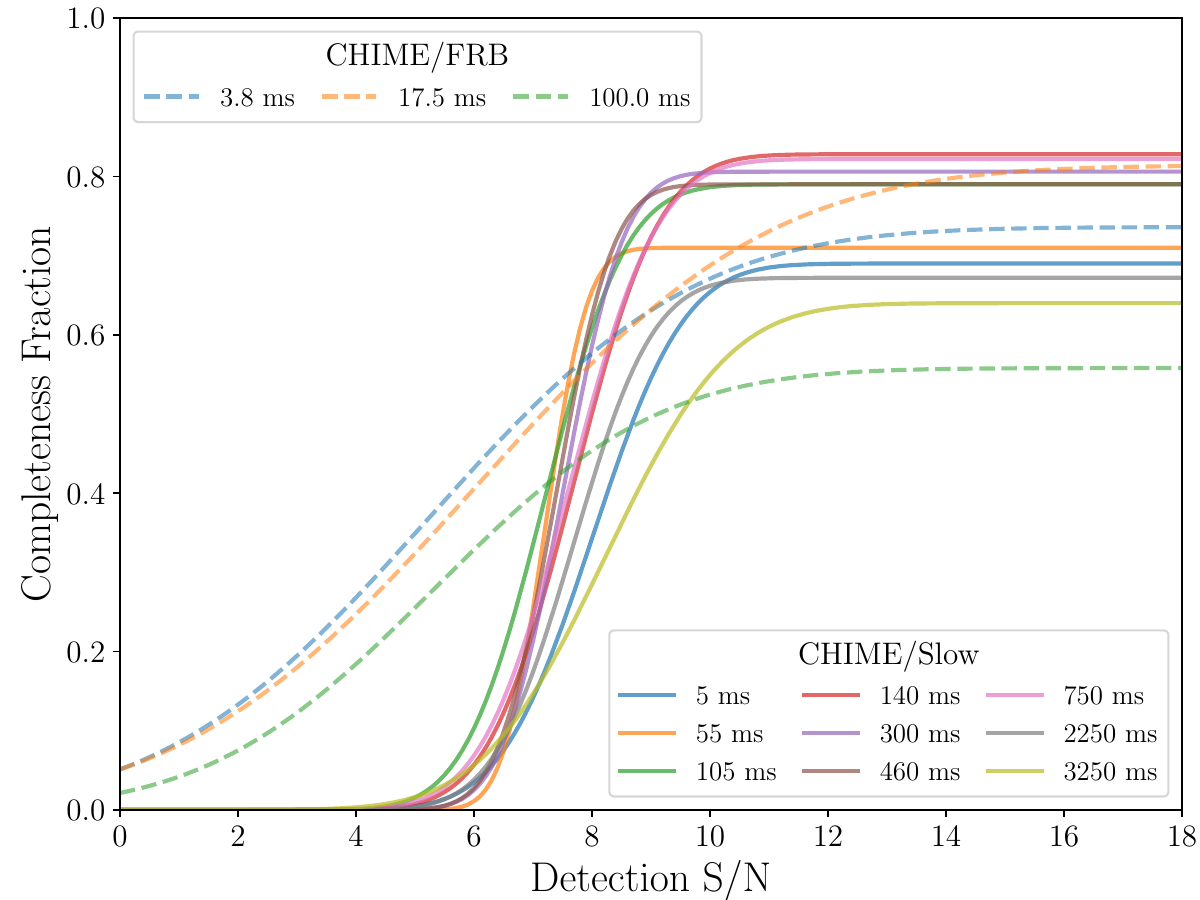}
\caption{\chimeslow and \chimefrb completeness as a function of detected S/N for different total (intrinsic + scattering) pulse widths. Only nine out of the 30 injected pulse width bins are plotted here. The first and the last bin in the each column of the legend approximately represent the narrowest and the widest pulse width bin injected in 16-ms, 128-ms and 512-ms data streams respectively.}\label{fig:completeness}
\end{figure}

Figure~\ref{fig:completeness} shows the completeness of \chimeslow and \chimefrb as a function of detected S/N for a few total injected pulse width bins. Note that the curves are smoothened by fitting a scaled version of the error function to the original data that was obtained following the method described above. It can be seen that \chimeslow performs better in recovering wider bursts. Note that the first \chimefrb injection run carried out as part of Catalog 1 analysis did not include pulses with intrinsic width~$>$100~ms.
\centerwidetable
\begin{rotatetable*}
\begin{deluxetable*}{ccccccccccc}
    \tablecaption{Properties of the bursts detected by CHIME/Slow. The values are estimated using the \texttt{fitburst} package and the daily calibration matrices stored by CHIME/FRB. The redshift for FRB~20230204C is estimated using the Macquart relation, for FRB~20220912A bursts it is taken from~\citet{Ravi2023}. It should be noted that the flux, fluence and luminosity values reported here are a 3-$\sigma$ lower limits due to the calibration method used in the analysis.}\label{tab:burst_properties}
    \tablehead{
    TNS Name & Arrival Time \tablenotemark{a} & S/N & DM & $\mathrm{W_{50}}$ & $\tau$\tablenotemark{b} & $\mathrm{z}$ & $\mathrm{D_L}$ \tablenotemark{c} & Peak Flux & Fluence & Luminosity \\
    & (UTC) &  & (\dm) & ($\mathrm{ms}$) & ($\mathrm{ms}$) & & ($\mathrm{Mpc}$) & ($\mathrm{Jy}$) & ($\mathrm{Jy\,ms}$) & ($\times$10$^{40}$  erg s$^{-1}$)}
    %\tabletypesize{\scriptsize}
    \startdata
    FRB 20221229A & 2022-12-29T00:38:43.030 & 11.11 & 223.50 $\pm$ 0.31 & 11.59 $\pm$ 1.39 & 22.55 $\pm$ 2.52 & \multirow{8}{*}{0.0771$_{-0.0001}^{+0.0001}$} & \multirow{8}{*}{361.25$_{--0.49}^{+0.49}$} & $>$  1.46 & $>$ 37.51 & $>$   1.77 \\
    FRB 20221230A & 2022-12-30T00:27:01.496 & 12.73 & 221.12 $\pm$ 0.17 & 10.22 $\pm$ 0.85 & 9.59 $\pm$ 1.42 & & & $>$  1.72 & $>$ 30.43 & $>$   2.74 \\
    FRB 20230102B & 2023-01-02T00:22:45.569 & 17.88 & 223.88 $\pm$ 0.15 & 16.44 $\pm$ 0.66 & --  & & & $>$  1.21 & $>$ 21.04 & $>$   3.92 \\
    FRB 20230102C & 2023-01-02T00:23:01.217 & 14.77 & 220.50 $\pm$ 0.37 & 25.48 $\pm$ 3.16 & --  & & & $>$  0.57 & $>$ 18.88 & $>$   1.22 \\
    FRB 20230106D & 2023-01-06T23:58:02.322 & 12.85 & 222.62 $\pm$ 0.20 & 17.22 $\pm$ 0.71 & -- & & & $>$  1.05 & $>$ 20.06 & $>$   1.67 \\
    FRB 20230107A & 2023-01-07T00:06:08.711 & 14.58 & 219.38 $\pm$ 0.77 & 19.73 $\pm$ 3.46 & 75.06 $\pm$ 14.89 & & & $>$  0.62 & $>$ 25.45 & $>$   2.01 \\
    FRB 20230107B & 2023-01-07T00:08:04.500 & 15.93 & 222.88 $\pm$ 0.21 & 16.08 $\pm$ 1.27 & 11.38 $\pm$ 1.89 &  & & $>$  1.31 & $>$ 25.36 & $>$   2.09 \\
    FRB 20230107C & 2023-01-07T00:10:48.098 & 11.48 & 223.38 $\pm$ 0.55 & 24.47 $\pm$ 1.70 & -- & & & $>$  0.76 & $>$ 23.56 & $>$   0.92 \\
    FRB 20230204C\tablenotemark{d} & 2023-02-04T02:41:04.028 & 16.59 & 386.25 $\pm$ 0.11 & 3.49 $\pm$ 1.32 & 98.31 $\pm$ 8.33 & 0.1500$_{-0.0800}^{+0.0800}$ & 736.92$_{--410.54}^{+447.10}$ & $>$  0.57 & $>$ 16.82 & $>$   9.57 \\
    \enddata
\tablenotetext{a}{The arrival time is topcentric and at the bottom of of observation band (400~MHz).}
\tablenotetext{b}{Scattering timescale at 400~MHz. A `--' indicates that adding a scattering tail did not improve the fit.}
\tablenotetext{c}{Luminosity distance is estimated assuming a flat-$\Lambda$CDM cosmology with the cosmological parameters taken from ~\citet{Planck2018VI}.}
\tablenotetext{d}{The FRB 20230204C is localized to RA = $53.02^\circ \pm 0.48^\circ$ and Dec = $50.01^\circ \pm 0.48^\circ$ by \chimefrb pipeline.}
\end{deluxetable*}
\end{rotatetable*}
\begin{deluxetable*}{ccccc}[h]
    \tablecaption{Injection parameters used to estimate \chimeslow pipeline completeness. The S/N values are sampled from a power-law distribution with an index of -1.5. All other values are sampled from a uniform distribution.}\label{tab:injection_params}
    \tablehead{
    Time Sampling & DM\tablenotemark{a} & Pulse Width (intrinsic) & S/N\tablenotemark{b}& Scattering Measure @ 1 GHz\\
    (ms) & (\dm) & (ms) & & (ms)
    }
    \startdata
    16& \multirow{3}{*}{100 -- 3000} & 1 -- 128 & 2.12 -- 2376.15 &\multirow{3}{*}{0.001 -- 5}\\
    128& & 120 -- 512 & 4.52 -- 4776.50 &\\
    512& & 500 -- 4096 & 4.50 -- 2016.92 &
    \enddata
\tablenotetext{a}{We choose 100~\dm as a lower bound for DM as at low DMs we are sensitive to Galactic pulsars. This could cause confusion between a real source and an injected source and we wanted to minimize this effect.}
\tablenotetext{b}{The lower bound of the injected S/N was chosen to be lower than the detection threshold of 10 to account for the fact that the definition of \texttt{simpulse} S/N (used for injection) and \texttt{PRESTO} S/N (used for detection) are not the same. The expected S/N is also affected by the pulse width, scattering measure and RFI enivronment. Hence we sample fainter S/N to ensure we probe the complete parameter space.}
\end{deluxetable*}
\section{Estimation of all-sky rate}\label{sec:rate}
We use the results from the pilot survey and the completeness analysis to estimate the all-sky rate of radio transients between 16~ms and 5~s. To determine the all-sky rate, we consider the one-off event FRB~20230204C as the only detection and discard the bursts from the repeating FRB~20220912A, as it was hyper-active during the pilot survey and contributed significantly to the all-sky rate of FRBs as shown by~\citet{Ould-Boukattine2025}.

The detected rate of transients depends on the intrinsic population model, the sky exposure, and the completeness of the detection pipeline. The intrinsic population model can be a function of multiple parameters (e.g., fluence, width, DM, scattering timescale etc.), but since we have only one detection, we assume that the intrinsic model only depends on fluence and width, and ignore the dependence on other parameters. Also, we consider width as the broadened width, which includes scattering, as completeness analysis was carried out on limited data, making it difficult to disentangle the two effects and constraining our analysis to consider broadened width only. Following this assumption, the observed rate ($N^d$) of transients between widths $w_\mathrm{min}$ and $w_\mathrm{max}$ and fluence $F_\mathrm{min}$, and between $F_\mathrm{max}$, is given by:
\begin{equation}\label{eq:rate_obs}
    N^d  = \int_{w_\mathrm{min}}^{w_\mathrm{max}} \int_{F_\mathrm{min}}^{F_\mathrm{max}} C(w) E(S) R(F, w; R_0, \beta) \mathrm{d}w \mathrm{d}F,
\end{equation}
where $C(w)$ is the completeness of the pipeline as a function of the width, $E(S)$ is the exposure (sky area $\times$ observing time) of the \chimeslow data, where the CHIME telescope has a relative sensitivity (with respect to the peak sensitivity under nominal operations and in minimal RFI environment) higher than $S$, and $R(F, w; R_0, \beta)$ is the intrinsic population model given by: 
\begin{equation}
    R (F,w;R_0, \beta) = \frac{-\alpha R_0}{F_0 w_0} \left(\frac{F}{F_0} \right)^{\alpha - 1}\left(\frac{w}{w_0}\right)^\beta,
\end{equation}
where, $R_0$ is the all-sky rate (sky$^{-1}$ day$^{-1}$) integrated over all fluence and width values, $F_0$ and $w_0$ are pivot values for fluence and width respectively and $\alpha$ and $\beta$ are the power-law indices for fluence and width dependence respectively. The functional form of all-sky rate dependence on fluence is fixed following CHIME/FRB~catalog~1~\citep{CHIMEFRB2021} with the assumption that the population is uniformly distributed in a flat Universe, setting the value of $\alpha$ to be --1.5.

For the survey, we had kept our detection threshold at S/N = 10, and hence we compute the completeness and sensitivity for this threshold. We consider a constant value of 0.8 for completeness fraction as from Figure~\ref{fig:completeness}, it can be seen that for most pulse width bins, the completeness is close to this value\footnote{Note that the injection analysis was done on a limited amount of data and hence we have not truly sampled the intrinsic noise properties. Therefore taking a constant and more conservative value of 0.8 for the completeness fraction is a safe assumption, as it will only lead to an underprediction of the all-sky rate.}. The sensitivity $S$ is a function of S/N threshold, fluence, width and also depends on the beam model and daily sensitivity variations. Details about how the sensitivity and exposure are estimated are given in Appendix~\ref{sec:appendixA}. Here we note that we consider 10\% of the peak sensitivity\footnote{The peak sensitivity is attained at the zenith of the beam.} of all beams in our data averaged over frequency, as the cutoff sensitivity and the sky area above this sensitivity is considered for exposure calculation.

We fit the observed rate of one detection in the entire 16~ms to 5~s pulse width range (considered as one bin) with the expected rate given by the Equation~\ref{eq:rate_obs} using Maximum Likelihood Estimation (MLE) to obtain the all-sky rate $R_0$. We use Binned Poisson likelihood~\citep{Zyla2020} for the MLE estimation, which is given as
\begin{equation}
    \ln \mathcal{L} = -\sum_i^M\left(N^e_i - N^d_i + N^d_i \ln \left[\frac{N^d_i}{N^e_i} \right]\right),
\end{equation}
where $M$ are the number of bins (one), and $N^e_i$ and $N^d_i$ are the expected and observed number of detections in each bin, respectively. The model is fit using MCMC analysis to obtain the posterior distribution of $R_0$. 

For the fitting, we assume a uniform prior in log$_{10}$ space for $R_0$, and use three different values of $\beta$ (--1, --2 and --3) to test the dependence of the rate with respect to three different population models for width. The minimum fluence value is set to 5~Jy~ms\footnote{This value is chosen to match the \chimefrb Catalog-1 fluence threshold~\citep{CHIMEFRB2021}.} and the maximum fluence value is set to 100~Jy~ms\footnote{This is chosen to be arbitrary high such that a burst of this fluence will be detectable by CHIME at a very low sensitivity as well.}. The parameters $F_0$ and $w_0$ in Equation~\ref{eq:rate_obs} are set to 5~Jy~ms and 16~ms, respectively.

\begin{figure}
    \centering
    \includegraphics[width=\linewidth]{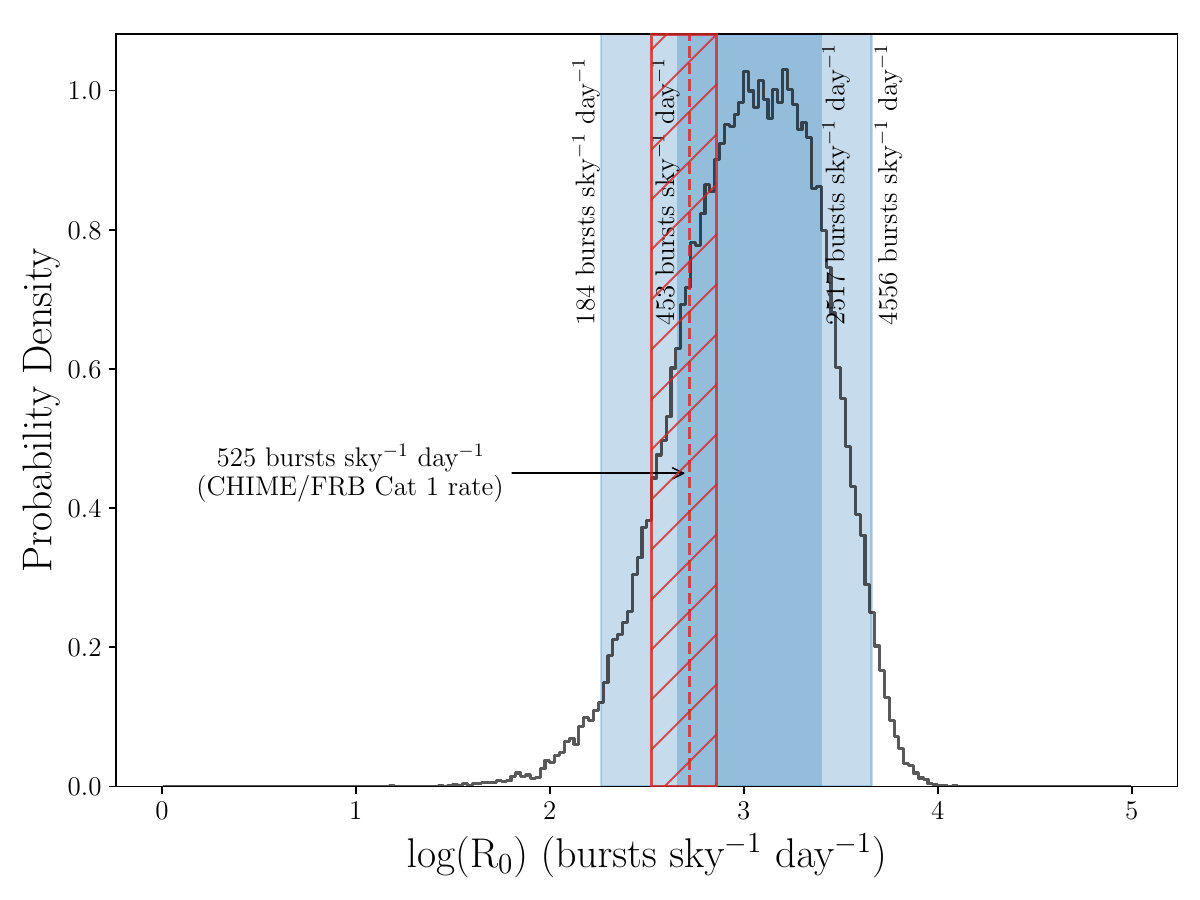}
    \caption{Posterior distribution of the all-sky rate $R_0$ in log space, marginalized over three different $\beta$ values (-1, -2 and -3). The light blue and dark blue shaded regions show the 95\% and 68\% credible interval respectively. The red vertical dashed line shows the all-sky rate of FRBs estimated by CHIME/FRB~Catalog~1~\citep{CHIMEFRB2021} for bursts with fluence above 5~Jy~ms and scattering time below 10~ms and the hatched region mark the 95\% credible interval for the \chimefrb rate.}\label{fig:rate_posterior}
\end{figure}

We use the \texttt{emcee}~\citep{Foreman-Mackey2013} package for the MCMC analysis. We marginalize the obtained $R_0$ posteriors over the three $\beta$ values. The resulting posterior distribution is shown in Figure~\ref{fig:rate_posterior}. As the rate estimation is done on a limited data set and we only have one detection, we do not report the mean all-sky rate, but instead report the lower and upper limit of the 95\% credible region. The estimated all-sky rate of radio transients having widths between 16~ms to 5~s, at observing frequency of 600 MHz and above the fluence of 5~Jy~ms is between $184$ to $4556$~bursts~sky$^{-1}$~day$^{-1}$ with 95\% confidence.

\section{Discussion}~\label{sec:disc}
The results of the pilot survey demonstrate the capability of \chimeslow to detect intrinsically wide as well as highly scattered radio transients and highlight the potential to discover new radio transients in the unexplored parameter space of few hundreds of milliseconds to few seconds. As the pilot survey was a proof-of-concept that was only later converted into a real-time pipeline, there is significant scope for modifications in the pipeline described here to improve the sensitivity and optimize the processing speed. Some of these improvements include: better RFI mitigation strategies (e.g., subtracting the average over neighboring beams; Kuiper et al. in prep.), optimization of the clustering parameters and multi-beam clustering to reduce false positives, retraining \texttt{FETCH} on the already collected \chimeslow data to improve the classification accuracy, improvements to the code efficiency to reduce the processing time and more systematic injection analyses to better characterize the sensitivity and completeness of the pipeline.

These optimization efforts are currently underway in parallel to running the current version of the pipeline as a real-time backend at the CHIME site with partial operations\footnote{We are searching about $\sim$300 beams out of the 1024 \chimefrb beams since Oct 2025.}. When fully operational, the real-time \chimeslow backend will be capable of searching the entire CHIME sky in real-time for radio transients between 16~ms to 5~s pulse widths, with a low-latency alert system to report detected transients to the community~\citep[e.g. via the \texttt{frb-voe} service,][]{Abbott2025}. Description of the real-time \chimeslow backend and results from the real-time survey will be presented in a future work.
\begin{figure}
    \centering
    \includegraphics[width=\linewidth]{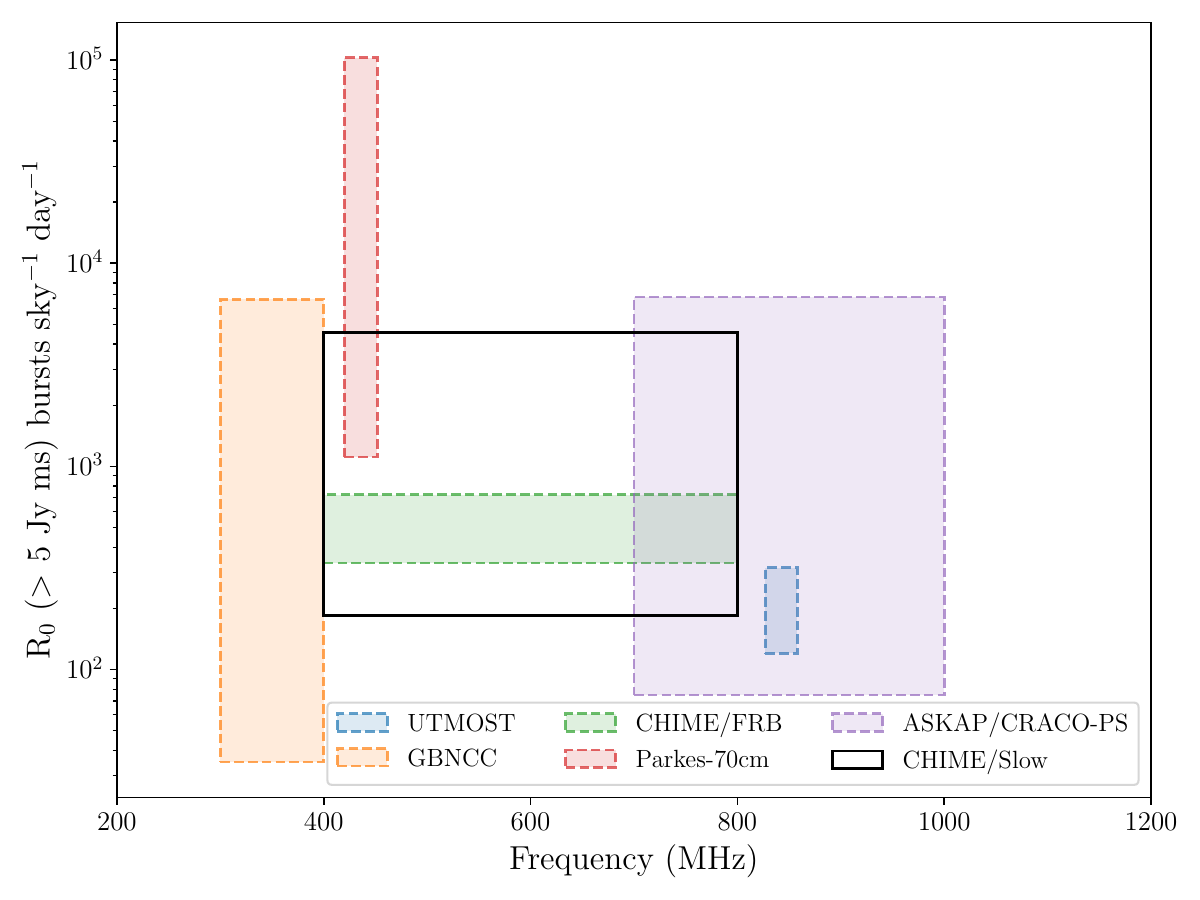}
    \caption{\chimeslow pilot survey all-sky rate of radio transients plotted along with the all-sky rate estimated by other surveys with center frequencies close to CHIME. All the rates are scaled to the \chimeslow and \chimefrb fluence threshold of 5 Jy ms assuming a Euclidean distribution of sources (i.e. $N (> F) \propto F^{-1.5} $). The data are taken from~\citet{Farah2019}(UTMOST),~\citet{Parent2020}(GBNCC),~\citet{CHIMEFRB2021}(\chimefrb),~\cite{Crawford2022,Wang2025}(Parkes-70cm),~\citet{Wang2025}(ASKAP/CRACO). Note that the Parks-70cm value reported here is the re-estimation of the original rate by~\citet{Wang2025} and not the rate reported by~\cite{Crawford2022}.}\label{fig:all_sky_rate_comparison}
\end{figure}

Figure~\ref{fig:all_sky_rate_comparison} shows the estimated all-sky rate from the \chimeslow pilot survey along with the all-sky rates estimated by other surveys at similar frequencies. From the figure, it can be seen that the \chimeslow pilot survey rate estimate is comparable to the rates estimated by other surveys, except for the upper limit of Parkes-70cm survey~\citep{Crawford2022} which is an order of magnitude higher than any other estimates. This value is a re-estimation of the original rate by~\citet{Wang2025} by considering only one out of four detections reported by~\citet{Crawford2022} to be of astrophysical origin\footnote{Two of the remaining three detections have low S/N and one has a narrow bandwidth in an already band-limited data (total bandwidth of 32 MHz) making the candidates marginal.}. The original rate reported by~\citet{Crawford2022} is significantly higher and is inconsistent with other survey rates. Consistent estimates on all-sky rates between \chimeslow and other surveys, particularly the ASKAP/CRACO-PS and Parkes-70cm in which pulse widths up to $\sim$200~ms were searched, clearly highlight the prediction that most current surveys, tuned mostly for millisecond-duration FRBs, could be missing a population of intrinsically wide and/or highly scattered FRBs. Hence, in addition to \chimeslow, the ongoing efforts by many other telescopes to search for wider duration radio transients are well placed to uncover this missing population.

One caveat to note while comparing the rates from different surveys with the rate predicted by \chimeslow pilot survey is that the maximum duration and DM searched in these surveys are different compared to our survey. \chimeslow pilot survey searched up to 5-s pulse widths and up to DM of 3000~\dm, while the maximum pulse width searched in other surveys is $\sim$200~ms; however, other surveys have searched much larger DMs (up to $\sim$13000~\dm in case of \chimefrb and ASKAP/CRACO-PS). Although, we do not expect our rate estimate to change significantly if we had searched for higher DMs as preliminary checks with injection data did not show lack of completeness at high DMs. To quantify the dependence of the rate on DM and on pulse widths, particularly $\gtrsim$ 1~s, more detections are necessary. Future work with a larger sample of detected transients from the real-time \chimeslow survey will help to better understand the pulse width and DM dependence and constrain the all-sky rate more accurately.

When comparing the \chimeslow pilot survey results with \chimefrb Catalog~1~\citep{CHIMEFRB2021} and injection system results~\citep{Merryfield2023}, the prediction by both works that \chimefrb could be missing highly scattered bursts is strongly supported by this work. The pilot survey has already detected one of the most highly scattered bursts by the CHIME telescope, and the all-sky rate predicted by the pilot survey is consistent with the \chimefrb all-sky rate supporting the hypothesis that there is an undetected population of highly scattered events in the \chimefrb data. The \chimeslow real-time pipeline will detect more such highly scattered bursts in the future which will be crucial to understand the local burst environments. These predictions, the results and the lessons learnt from the systematic \chimeslow real-time survey will be crucial to guide any future upgrades to the CHIME telescope to incorporate wider pulse width searches and also to guide any future surveys.

\section{Conclusions}~\label{sec:concl}
The discovery of FRBs prompted the development of new search techniques to systematically search for FRB-like radio transients. However, due to some technological limitations (e.g. RFI mitigation algorithms) these searches have been limited to millisecond-duration transients, and wider pulse widths have been largely unexplored. Recent discoveries of long-period radio transients, as well as some theoretical predictions of FRB-like radio transients associated with compact binary mergers and GRBs, highlight the potential of a systematic search for second-duration radio transients. The \chimeslow backend attempts to fill this unexplored parameter space by systematically searching for radio transients with 16~ms to 5~s pulse widths. 

In this article, we have described the prototype version of the \chimeslow pipeline and results from the \chimeslow pilot survey carried out using archival data. The \chimeslow pipeline is based on well-tested tools such as \presto, \hdbscan, and \fetch and has a built-in injection routine to benchmark its performance. The pipeline was tested on archival data collected in December 2022 and January 2023 and February 2023. In these data, nine bursts were detected, one of which is a new non-repeating FRB and eight are from a hyperactive known repeater. We also report the lower and upper limit (95\% credible region) on the all-sky rate of radio transients with pulse widths between 16~ms to 5~s, at observing frequency of 600 MHz, and fluence threshold of 5~Jy~ms to be $184$ and $4556$ ~bursts~sky$^{-1}$~day$^{-1}$

The \chimeslow pilot survey pipeline is being ported to run as a real-time backend at the CHIME site. The backend is already partially online in its current version, and efforts are underway to optimize the pipeline for real-time operations and to scale the system to handle the full CHIME sky. We expect to have the real-time \chimeslow backend fully operational by mid to late 2026.

%% Also note that the akcnowlodgment environment does not support long amounts of text. If you have a lot of people and institutions to acknowledge, do not use this command. Instead, create a new \section{Acknowledgments}.
% \begin{acknowledgments}
\section*{Acknowledgments}
We acknowledge that CHIME is located on the traditional, ancestral, and unceded territory of the Syilx/Okanagan people. We are grateful to the staff of the Dominion Radio Astrophysical Observatory, which is operated by the National Research Council of Canada. CHIME is funded by a grant from the Canada Foundation for Innovation (CFI) 2012 Leading Edge Fund (Project 31170) and by contributions from the provinces of British Columbia, Québec and Ontario. The \chimefrb Project is funded by a grant from the CFI 2015 Innovation Fund (Project 33213) and by contributions from the provinces of British Columbia and Québec, and by the Dunlap Institute for Astronomy and Astrophysics at the University of Toronto. 

CHIME/Slow is supported by Dunlap seed funding provided by the Dunlap Institute. The Dunlap Institute is funded through an endowment established by the David Dunlap family and the University of Toronto.

We thank Parag Shah from TIFR for the help in setting up the compute on which the pilot survey was carry out. We are grateful to Scott Ransom for his inputs on configuring \presto to make it compatible with \chimeslow data. We thank Natasha Hurley-Walker, Manisha Caleb and Iris de Ruiter for sharing the data for Figure~\ref{fig:transient_ps} and Andy Wang for sharing the data for Figure~\ref{fig:all_sky_rate_comparison}.

Z.P. is supported by an NWO Veni fellowship (VI.Veni.222.295). P.S. acknowledges the support of an NSERC Discovery Grant (RGPIN-2024-06266). F.A.D was supported by a NRAO Jansky Fellowship, F.A.D is supported by an NRC SKA Fellowship. The AstroFlash research group at McGill University, University of Amsterdam, ASTRON, and JIVE is supported by: a Canada Excellence Research Chair in Transient Astrophysics (CERC-2022-00009); an Advanced Grant from the European Research Council (ERC) under the European Union's Horizon 2020 research and innovation programme (`EuroFlash'; Grant agreement No. 101098079); and an NWO-Vici grant (`AstroFlash'; VI.C.192.045). V.M.K. holds the Lorne Trottier Chair in Astrophysics \& Cosmology, a Distinguished James McGill Professorship, and receives support from an NSERC Discovery grant (RGPIN 228738-13). K.N. acknowledges support by NASA through the NASA Hubble Fellowship grant \# HST-HF2-51582.001-A awarded by the Space Telescope Science Institute, which is operated by the Association of Universities for Research in Astronomy, Incorporated, under NASA contract NAS5-26555. FRB and Pulsar research at UBC is supported by an NSERC Discovery Grant and by the Canadian Institute for Advanced Research.
% \end{acknowledgments}


%\vspace{5mm}
\facility{CHIME/FRB}


\software{NumPy \citep{numpy2020}, SciPy \citep{scipy2020}, Matplotlib \citep{matplotlib2007}, Astropy \citep{astropy2022}, PRESTO \citep{Ransom2011}, HDBSCAN \citep{McInnes2017}, FETCH \citep{Agarwal2020}, emcee \citep{Foreman-Mackey2013}}

\appendix
\section{Exposure Calculation}\label{sec:appendixA}
Here we explain the details of the exposure calculation necessary to forward-model the all-sky rate following equation~\ref{eq:rate_obs}. The exposure depends on the CHIME/FRB beam model, the daily sensitivity variations, and the daily observing time. First, we define a relative sensitivity metric ($S$), for a given intrinsic width ($w$) and fluence ($F$), as the minimum sensitivity required to detect a burst above the S/N threshold of $\mathrm{S/N_{thr}}$ = 10. This sensitivity is defined as:
\begin{equation}
    S = \frac{\mathrm{S/N}_\mathrm{thr}}{\mathrm{S/N}_\mathrm{max}},
\end{equation}
where $\mathrm{S/N_{max}}$ is the maximum S/N of a burst with fluence $F$ and width $w$ at the peak sensitivity of the CHIME beam (i.e. at zenith, under nominal operations with no excess RFI). The $\mathrm{S/N_{max}}$ is computed using the radiometer equation following~\citet{Chawla2017} as:
\begin{equation}
    \mathrm{S/N_{max}} = \frac{fGw}{\beta (T_{sys} + T_{sky})} \sqrt{\frac{n_p \Delta \nu}{w_b}},
\end{equation}
where $f$ is the flux density of the burst ($F/w_b$), $G$ is the telescope gain at zenith, $w$ is the intrinsic width (including scattering) of the burst, $\beta$ is a factor to account for the digitization loss, $T_{sys}$ and $T_{sky}$ are the system and sky temperatures respectively, $n_p$ is the number of polarizations summed, $\Delta \nu$ is the bandwidth, and $w_b$ is the broadened width. Since we do not differentiate between the intrinsic width and scattering in our analysis, the definition of broadened width is given by:
\begin{equation}
    w_b = \sqrt{w_i^2 + t_{samp}^2 + t_{DM}^2},
\end{equation}
where $t_{samp}$ is the sampling time, $t_{DM}$ is the intra-channel DM smearing time. 

We use typical values from CHIME/FRB for the parameters in the radiometer equation, with $G = 1.38$~K~Jy$^{-1}$, $\beta = 1.0, T_{sys} = 50$~K, $n_p = 2$, and $\Delta \nu = 200$~MHz (accounting for about 50\% loss in the number of channels due to the RFI masking). The sky temperature, $T_{sky} = 10$~K, is estimated by extrapolating (with spectral index of --2.6) the reprocessed Haslam 408~MHz sky temperature map~\citep{Remazeilles2015} to the CHIME central frequency of 600~MHz, and taking the average value over the observed part of the sky. When computing the broadened width and intrinsic width, as mentioned in Section~\ref{sec:rate}, we consider $w$ to include scattering effects, and hence the broadened width only includes the sampling time as we ignore the intra-channel DM smearing, since contribution from it is minimal at typical DMs.

\begin{figure}[!h]
    \centering
    \includegraphics[width=0.5\linewidth]{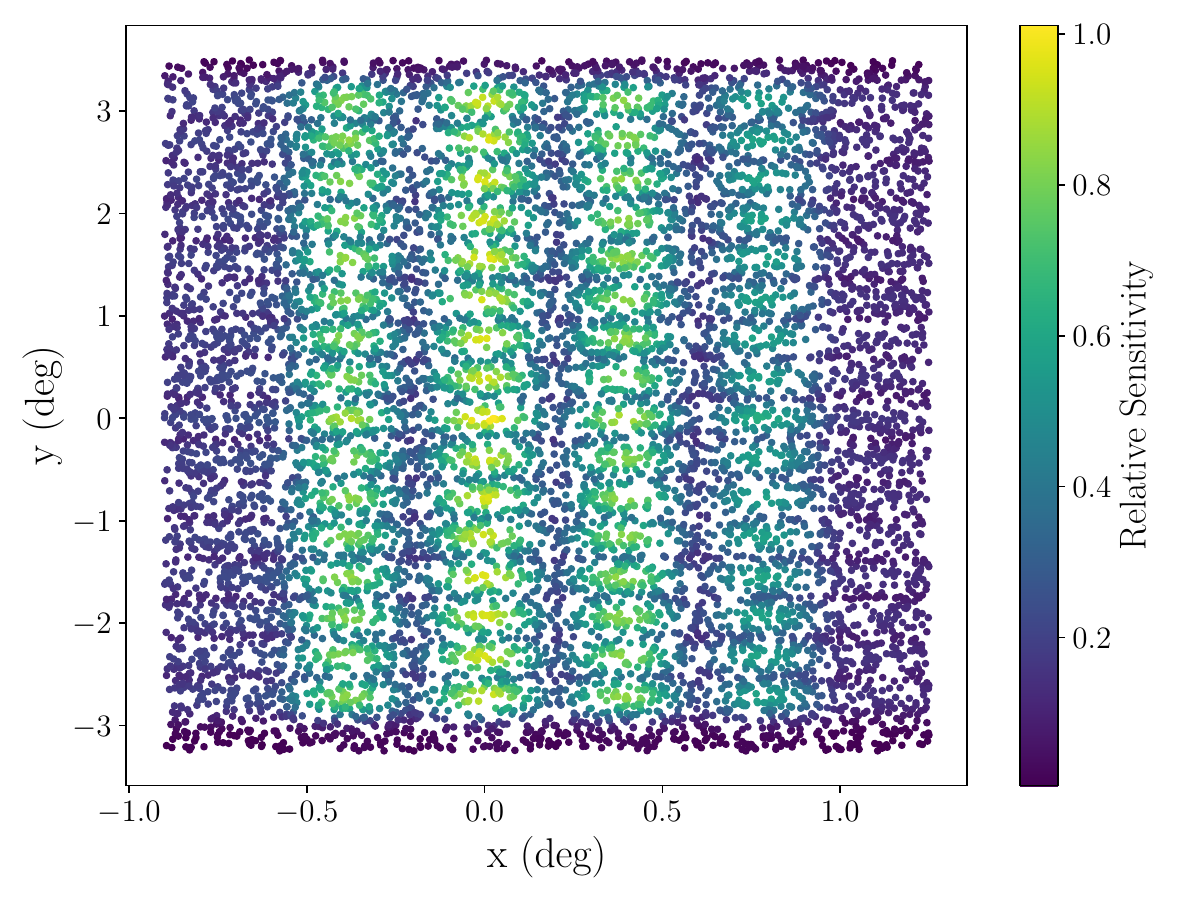}
    \caption{Relative sensitivity as a function of sky position for all 64 beams in the CHIME/Slow data. The X and Y axes represent angles in telescope coordinates, and are aligned in the North--South and East--West directions respectively. }\label{fig:beam_sensitivity}
\end{figure}

As mentioned in Section~\ref{sec:rate}, the exposure ($E(S)$, sky area $\times$ observing time, see Equation~\ref{eq:rate_obs}) is estimated assuming a relative sensitivity of 10\% of the peak sensitivity. Assuming this, we first estimate the cumulative sky fraction visible above this threshold sensitivity to all 64 beams for which the \chimeslow pilot survey data was collected. The \chimefrb beam model gives the sensitivity of each beam for a point on the sky relative to the zenith as a function of frequency\footnote{For details regarding \chimefrb beams and their orientation, see~\citet{CHIMEFRB2021}}. To estimate the cumulative sky fraction, we first randomly sample the sky in the East-West and North-South directions centered at telescope meridian, and only covering the sky area corresponding to the 16 beam rows, and find frequency averaged sensitivity for each beam at each position using the beam model. Then for each position in the sky, we take the maximum of the sensitivities over the beams, and consider that value as the sensitivity of the telescope at that position (see Figure~\ref{fig:beam_sensitivity}). 

\begin{figure}[h]
    \centering
    \includegraphics[width=\linewidth]{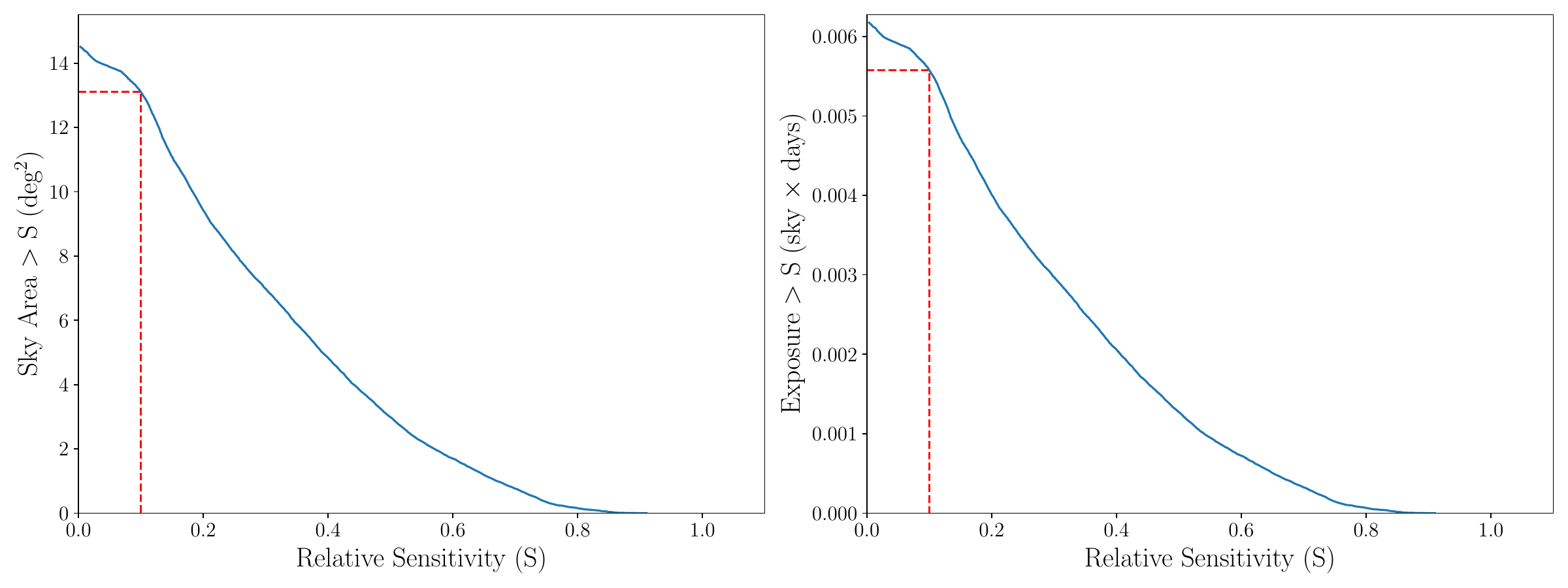}
    \caption{\textbf{Left}: Sky area (in deg$^2$) above the relative sensitivity of $S$. The horizontal and vertical lines mark the area above the 10\% relative sensitivity threshold used for the exposure calculation. This area corresponds to 13.10 deg$^2$. \textbf{Right}: The total exposure (sky $\times$ days) above the relative sensitivity $S$. The horizontal and vertical lines mark the total exposure above the 10\% relative sensitivity threshold.}\label{fig:sky_area_exp}
\end{figure}

We then assume that the total area of each sampled point is $d$A, which is given by A/N, where A is the total area sampled and N is the number of points sampled. To calculate the cumulative sky fraction above 10\% relative sensitivity, we sort the sensitivities and take the cumulative sum of the area. This is further corrected for the daily sensitivity variations by multiplying the sensitivities with the daily sensitivity factor, and then finding the sky area corresponding to the corrected 10\% sensitivity. The daily sensitivity factor is estimated based on the total number of pulsar detected by CHIME/FRB in a day compared with expected number of pulsar detections under nominal operating conditions. The corrected cumulative sky area above the relative sensitivity of 10\% is shown in Figure~\ref{fig:sky_area_exp}~(Left). This area is then converted to the daily exposure by finding the fraction of the sky that is visible per day, and multiplying that with the total number of observing days. Figure~\ref{fig:sky_area_exp}~(Right) shows the exposure as a function of relative sensitivity. The exposure is then used in Equation~\ref{eq:rate_obs} to estimate the observed rate of transients.




\bibliography{chimeslow.bib}
\bibliographystyle{aasjournal}



\end{document}